\documentclass[a4paper, twoside]{article}
\usepackage[ngerman]{babel}
\usepackage[T1]{fontenc}			%
\usepackage[latin1]{inputenc} % windows characters
\usepackage{fancyhdr}
\usepackage{graphicx}					% including graphic
\usepackage{amsmath, amsthm, amssymb}	%math symbols
\usepackage{icomma}						% for correct deciaml separator

\usepackage{caption} 
\DeclareCaptionLabelSeparator{space3}{\;\;\;}
\usepackage{times}
\fontfamily{ptm}\selectfont  % times family

\renewcommand{\small}        {\fontseries{m}\fontsize{10}{11}\selectfont}
\renewcommand{\normalsize}   {\fontseries{m}\fontsize{12}{15}\selectfont}

\renewcommand{\Large}        {\fontseries{m}\fontsize{14}{16}\selectfont}
\renewcommand{\LARGE}        {\fontseries{m}\fontsize{18}{22}\selectfont}
\newcommand{\refTab}[1]{Tabelle~\ref{#1}}
\newcommand{\refFig}[1]{Abb.~\ref{#1}}

\begin{document}

\LARGE{\bf Big Data: Perspektiven für Smart Grids und Smart \mbox{Buildings}}
\Large 

R. Mikut\\
\normalsize
{\it Institut für Angewandte Informatik}\\
{\it Karlsruher Institut für Technologie}\\
{\it Hermann-von-Helmholtz-Platz 1, 76344 Eggenstein-Leopoldshafen
}\\
{\it EMail: ralf.mikut@kit.edu }

%#################################################################################################
\normalsize
\vspace*{6mm}
{\bf Zusammenfassung}
{\small Dieser Beitrag gibt eine kurze Übersicht über aktuelle Trends auf dem Gebiet Big Data. Nach einer Klärung des Begriffs ''Big Data'' und einer Erläuterung der Entstehung großer Datenmengen werden anwendbare Methoden skizziert. Abschließend geht der Beitrag verstärkt auf gegenwärtig etablierte Anwendungen und zukünftige Anwendungspotenziale in den Gebieten intelligenter Gebäude (''Smart Buildings'') und intelligenter Netze (''Smart Grids'') ein.}

\vspace*{6mm}

{\bf Abstract} 
{\small This paper present a short survey about recent trends on the arising field of big data. After a definition and explanation of ''Big Data'' and a discussion why data sizes increase, appropriate methods to solve big data problems are introduced. In addition, recent applications and future potentials in smart buildings and smart grids are discussed.}
%#################################################################################################
%
%%%%%%%%%%%%%%%%%%%%%%%%%%%%%%%%%%%%%%%%%%%%%%%%%%%%%%%%%%%%%%%%%%%%%%
\section{Einführung}
%%%%%%%%%%%%%%%%%%%%%%%%%%%%%%%%%%%%%%%%%%%%%%%%%%%%%%%%%%%%%%%%%%%%%%
In den letzten Jahren hat auch in technischen Anwendungen eine Reihe von Themen große Aufmerksamkeit erlangt, die mit der Nutzung großer Datenmengen (''Big Data'') und der umfassenden Digitalisierung von industriellen Prozessen (''Industrie 4.0'') zu tun haben. 

Industrie 4.0 \cite{Bauernhansl14,Kagermann13} steht für die sogenannte vierte industrielle Revolution (nach der Dampfmaschine, der Elektrifizierung/Massenproduktion und der Computerisierung). Das Ziel sind veränderte Wertschöpfungsketten, die auf der Verfügbarkeit \emph{aller} relevanten Daten in Echtzeit und ihrer Vernetzung beruhen. Das schließt die Erfassung und Analyse heterogener Daten (Zeitreihen, Bilder, Videodaten, Audiodaten, Protokolldateien mit Events, Textdateien) und ihre Fusion ein. 

In diesem Beitrag werden Anwendungsszenarios für Gebäudedaten und für die intelligente Vernetzung von technischen Infrastrukturen näher betrachtet. Mögliche Datenquellen sind hier Sensoren, die örtlich und zeitlich aufgelöste Daten u.a. über mechanische Verformungen, Innen- und Außentemperaturen, Feuchtigkeit sowie elektrische Leistung, Warmwasser-, Kaltwasser- und Gasmengen generieren. Dazu kommen Protokolldateien mit erkannten Events, die Aufschluss über strukturelle Änderungen im jeweils betrachteten System sowie über erkannte Fehlersituationen geben. Solche Daten haben großes Potenzial zur Analyse der Sicherheit von Gebäuden und Infrastrukturen sowie zur Optimierung in Richtung Nutzerzufriedenheit und effizientem Betrieb. Das gilt sowohl für einzelne Gebäude als auch für vernetzte Infrastrukturen, z.B. Energienetze (Smart Grids, \cite{Fang12}) oder intelligente Städte (Smart Cities, \cite{Kitchin14}).     

Das Ziel dieses Beitrags besteht darin, 
\begin{itemize}
\item den Begriff Big Data näher zu erläutern (Abschnitt~\ref{sec:big_data}),
\item auf die zugrundeliegenden generischen Methoden einzugehen (Abschnitt~\ref{sec:methods}) und 
\item die Anwendungsszenarien  Smart Buildings (Abschnitt~\ref{sec:smart_buildings}) und Smart Grids (Abschnitt~\ref{sec:smart_grids}) näher zu erläutern.
\end{itemize}

%%%%%%%%%%%%%%%%%%%%%%%%%%%%%%%%%%%%%%%%%%%%%%%%%%%%%%%%%%%%%%%%%%%%%%
%
\section{Was bedeutet ''Big Data''?\label{sec:big_data}}
Der Begriff ''Big Data'' \cite{Chen14MNA,Wrobel14} hat seinen Ursprung eher in der Informatik und wird durch die folgenden Eigenschaften (''4Vs'') definiert \cite{Laney01,Ward13,Mikut16at}:
\begin{enumerate}
\item Volume (Größe bzw. Datenmenge), 
\item Velocity (erforderliche Geschwindigkeit beim Generieren, Einlesen, Auslesen oder Übertragen von Daten), 
\item Variety (Vielfältigkeit oder Heterogenität, bezogen auf viele verschiedene Datentypen und zu vernetzende Systeme mit unterschiedlichen Eigenschaften) und 
\item Veracity (Vertrauenswürdigkeit, bezogen auf den Anteil fehlerhafter, fehlender oder nur mit Unsicherheiten verfügbaren Daten wie z.B. Intervalldaten).
\end{enumerate}
Große Datenmengen resultieren entweder aus extrem vielen Datentupeln $N$ (z.B. bei Webzugriffen oder Protokolldateien mit Events) oder aus der Multidimensionalität, siehe \refTab{tab:dimension}. Diese Multidimensionalität tritt sowohl bei Zeitreihen (große Anzahl von Abtastzeitpunkten $K$ durch lange Aufzeichnungen und hohe zeitliche Auflösungen) als auch bei Bildern (viele Bildspalten und -zeilen $I_x,I_y$, evtl. mit einer zusätzlichen räumlichen Auflösung $I_z$) und Videos auf. In einigen Fällen kann auch die Anzahl $s$ der Einzelmerkmale, Zeitreihen und Bilder eine entscheidende Rolle spielen, z.B. bei vielen Sensoren. Hinzu kommt die Tatsache, dass sich verschiedene Arten der Dimensionalität multiplikativ verknüpfen, z.B. bei Zeitreihen mit $s \cdot K$. 
\begin{table}[hbt]
\centering
\caption{\label{tab:dimension}Datensätze für verschiedene Arten von Rohmerkmalen. Abkürzungen: $l = 1, \ldots, s$ Nummer gleichartiger Merkmale; $n=1,\ldots,N$ Datentupel; $k=1,\ldots,K$ Abtastzeitpunkte; ID: Eindeutige Identifier eines Events; $i_{E1}, \ldots, i_{Es}$ Parameter von Events; $i_x=1,\ldots, I_x$ Bildspalten; $i_y=1,\ldots,I_y$ Bildzeilen; $i_z=1,\ldots,I_z$ Bildschichten. Die niedrigeren Dimensionszahlen der Datensätze gelten für $s=1$ (nur ein Einzelmerkmal, eine Zeitreihe, ein Bild bzw. ein Video) sowie
ein Datentupel ($N=1$), modifiziert nach \cite{Mikut08}.}
\begin{tabular}{|p{0.22\textwidth}|c|p{0.3\textwidth}|p{0.25\textwidth}|}
\hline
 Rohdaten        &  Dimension & Bestandteile  & Anzahl Rohmerkmale $s_{Roh}$ pro Datentupel\\ \hline
 Einzelmerkmale  &  0-2       & $x_l[n]$ & $s $\\ \hline
 Zeitreihen      &  1-3       & $x_{ZR,l}[k,n]$ & $s \cdot K $\\ \hline
 Events          &  2         & $x_{Event}[ID, i_{E_1}, \ldots, i_{E_s},n]$ &  $E_s+1$ \\ \hline 
 Bilder          &  2-4       & $x_{Bild,l}[i_x,i_y,n]$  & $s \cdot I_x \cdot I_y $  \\ \hline
 3D-Bilder       &  3-5       & $x_{Bild,l}[i_x,i_y,i_z,n]$ & $s \cdot I_x \cdot I_y  \cdot I_z$ \\ \hline
 Videobilder     &  3-5       & $x_{Video,l}[i_x,i_y,k,n]$& $s \cdot I_x \cdot I_y \cdot K$    \\ \hline
 3D-Videobilder  &  4-6       & $x_{Video,l}[i_x,i_y,i_z,k,n]$&  $s \cdot I_x \cdot I_y \cdot I_z \cdot K$    \\ \hline
\end{tabular}
\end{table}

Big Data ist ein relativer Begriff: Probleme treten immer nur auf, wenn eine geplante oder bereits realisierte Anwendung in mindestens einem der obengenannten Punkte die verfügbare Software- und Hardwareinfrastruktur in einer Firma, einer Universität oder einer Forschungseinrichtung überfordert. Der Übergang zwischen klassischen Datenanalyseaufgaben und Big-Data-Aufgaben ist allerdings fließend, wenn Anwendungen an der Kapazitätsgrenze der bestehenden Infrastruktur liegen. Solche Aufgaben sind dann in der Regel mit klassischer Infrastruktur bei erhöhtem Aufwand und schlechter Performanz immer noch durchführbar, allerdings wäre eine spezielle, auf Big-Data-Aufgaben zugeschnittene Infrastruktur vorteilhaft. 

\section{Welche Methoden eignen sich im Umgang mit ''Big Data''?}\label{sec:methods}
Prinzipiell basieren alle Analysemethoden für Big-Data-Probleme auf etablierten Datenanalysemethoden, die seit Jahrzehnten erfolgreich eingesetzt werden. Diese Methoden stammen sowohl aus der klassischen Statistik \cite{Backhaus00,Jain00} als auch aus dem Maschinellen Lernen und der Computational Intelligence \cite{Kroll16,Michie94}. Beispiele für Verfahren des Maschinellen Lernens sind $k$-Nearest-Neighbor-Verfahren (Analyse der Ausgangsgröße ähnlicher Beispiele, \cite{Cover67}) und Support-Vektor-Maschinen (Suche optimaler Trennflächen in höherdimensionalen Räumen mit integrierten nichtlinearen Transformationen \cite{Burges98}). Zur Computational Intelligence gehören u.a. Künstliche Neuronale Netze \cite{Haykin09}, die biologisch inspirierte Verschaltungen einfacher Verarbeitungseinheiten zur Nachbildung nichtlinearer Zusammenhänge nutzen, und Cluster-Verfahren zur automatischen Suche nach ähnlichen Beispielen wie Fuzzy-C-Means \cite{Bezdek81,Hoeppner99}.   

Dennoch ergeben sich bei Big-Data-Aufgaben in einigen Punkten Besonderheiten:  
\begin{enumerate}
\item Es wird größerer Wert auf gut skalierende Algorithmen (z.B. mit geringerer Komplexität oder guter Parallelisierbarkeit) gelegt. Das ist insbesondere dann wichtig, wenn die Datenmengen schneller steigen als die Rechengeschwindigkeiten und Speicherkapazitäten der verfügbaren Infrastruktur. 
\item Hierfür sind teilweise neue Implementierungen erforderlich, die für verteilte Umgebungen konzipiert sind und gut skalieren. Beispiele sind verteilte Filesysteme wie Hadoop, Analysetools wie Apache Spark und Tools zum Cloud Computing, siehe z.B. \cite{Duepmeier15,Hashem15,Stegmaier16AT}, sowie spaltenorientierte Datenbanken \cite{Plattner14}.    
\item Zur Beherrschung multidimensionaler Daten werden geeignete Einzelmerkmale durch eine Merkmalsextraktion aus Zeitreihen und Bildern ermittelt, um Informationen in komprimierter Form enthalten. Auch diese Methoden sind seit langem bekannt, werden aber verstärkt eingesetzt. Darüber hinaus wird versucht, diesen Schritt durch integrierte Merkmalsextraktionen zu automatisieren, was oft unter dem Begriff Deep Learning zusammengefasst wird \cite{LeCun15}.  
\item Verfahren zur automatischen Beurteilung der Daten- und Modellqualität \cite{Batini09,Madnick09,Michie94,Pimentel14} spielen eine größere Rolle, weil die bisherigen halbautomatischen Techniken zur Datenbereinigung, Ausreißerdetektion und Modellvalidierung zu aufwändig werden und weil bestimmte Datenqualitätseffekte durch Standardvalidierungsmethoden nicht erkannt werden, siehe z.B. \cite{Doneit14}.
\item Zur Beherrschung von Datenströmen (also kontinuierlich anfallenden Daten, engl. data streams) müssen rekursive Methoden eingesetzt werden, die Modelle ständig nachführen und mit neu hinzukommenden Daten verbessern \cite{Aggarwal07}. 
\end{enumerate}

Eine Auswahl typischer Datenanalysemethoden zeigt~\refTab{tab:methoden}. Für eine erfolgreiche Lösung müssen Fachexperten für das jeweilige Anwendungsgebiet und Fachexperten für Datenanalyse eng zusammenarbeiten, um die Fragestellung in der Anwendung in ein für Datenanalysemethoden lösbares Problem zu übersetzen und um die Ergebnisse zu interpretieren.  Hierbei ist entscheidend, den kompletten Auswerteablauf durch geeignete Softwaretools \cite{Mikut11Wiley} zu unterstützen und skriptbasiert zu automatisieren. Dazu existieren kommerzielle Produkte wie der IBM SPSS Modeler (\emph{http://www.spss.com/software/modeling/modeler}) oder Open-Source-Lösungen wie KNIME (\emph{http://www.knime.org}, \cite{Berthold08}), Gait-CAD (\emph{http://sourceforge.net/projects/gait-cad}, \cite{Burmeister08}), sowie neuere Big-Data-Tools wie Apache Spark (\emph{http://spark.apache.org}, \cite{Meng16}).
\begin{table}[hbt]
	\centering		
		\caption{Ausgewählte Analysemethoden\label{tab:methoden}}
		\begin{tabular}{|p{0.33\textwidth}|p{0.65\textwidth}|}
		\hline
		Aufgaben                                    & Ausgewählte geeignete Verfahren		 \\ \hline
		Erkennung atypischer Werte (Ausreißer)      & Klassifikation ($k$-Nearest Neighbor, Support-Vektor-Maschinen)   \\ \hline
		Fehlerdetektion                             & Klassifikation (Bayes-Klassifikatoren, Künstliche Neuronale Netze, Support-Vektor-Maschinen) \\ \hline
		Prognose zukünftiger Werte einer Zeitreihe  & Regression (Autoregressive Modelle mit oder ohne externe Eingänge, 
		                                              Künstliche Neuronale Netze, $k$-Nearest Neighbor) \\ \hline
	  Suche nach Zusammenhängen in Protokolldateien mit Events  & Process Mining \cite{Aalst07} \\ \hline 
		Suche nach Subgruppen und typischen Mustern & Cluster-Verfahren (Fuzzy-C-Means, Hierarchisches Clustern), Spezielle Visualisierungen \\ \hline
		Verstehen wesentlicher Zusammenhänge        & Korrelationsanalysen, Regression (Polynome, Künstliche Neuronale Netze), Spezielle Visualisierungen \\ \hline		
		
		\end{tabular}
\end{table}

Typische klassische Datenanalyseaufgaben und typische Big-Data-Aufgaben zeigt \refTab{tab:anwendungen}. Auffällig sind dabei die extrem unterschiedlichen Anwendungsgebiete, die dennoch mit einem ähnlichen Satz an Methoden bearbeitet werden. 
\begin{table}[hbt]
	\centering		
		\caption{Typische klassische Datenanalyseaufgaben und Big-Data-Aufgaben\label{tab:anwendungen}}
		\begin{tabular}{|p{0.2\textwidth}|p{0.35\textwidth}|p{0.35\textwidth}|}
		\hline
		Aufgaben   & Klassische Datenanalyseaufgaben & Big-Data-Aufgaben 		 \\ \hline
	  Bauwesen   & Bruchortung \cite{Holst07}, Strukturanalysen mit zeitabhängigem Materialverhalten \cite{Graf12} & Echtzeitmonitoring-System für Gebäude \cite{Park13} \\ \hline
		Biologie   & Herzschlagerkennung aus kurzen Videos \cite{Pylatiuk14} & Analyse von Gensequenzen \cite{Armant13,Venter01}, 3D-Videobilder in der Mikroskopie \cite{Ahrens13,Kobitski15,Stegmaier16}\\ \hline
		Chemie     & Entwurf antibakterieller Peptide \cite{Mikut09} & 3D-Strukturerkennung von Proteinen \cite{Groom14}\\ \hline
		Ingenieurwesen & Greifplanung \cite{Beck03CDC,Bohg14}, Prognose für Photovoltaik-Anlagen \cite{GonzalezOrdiano16} und Windanlagen \cite{Bremnes04} & (Prädiktive) Fernwartung von Werkzeugmaschinen \cite{Bauer16}, Autonomes Fahren \cite{Werling15}, Cloud Robotics  \cite{Kehoe15},  Humanoide Robotik \cite{Asfour13}, Smart Meter Daten \cite{Waczowicz14_Eng}, Phasordaten \cite{Maass15} \\ \hline
	  Medizin    & Bewegungsanalysen \cite{Wolf14}, Bildverarbeitung \cite{Naik08}, Brain-Machine-Interfaces \cite{Mikut08IEEE,Millan10}, Statistische Analysen \cite{Wagner11}, Rollstuhlsteuerung \cite{Schmalfuss15} & Personalisierte  Medizin \cite{Hamburg10} \\ \hline  
		Physik     & -- & Elementarteilchenerkennung im Large Hadron Collider \cite{Bird11}, Klimaforschung \cite{Faghmous14}\\ \hline
		Sicherheits\-technologien   & Betrugserkennung \cite{Phua04} & Überwachung \cite{Lyon14}  \\ \hline
		Wirtschafts\-wissenschaften & Kundenbindungsanalysen \cite{Smith00} & Analyse sozialer Medien \cite{Tufekci14}, Logistik \cite{Kovynyov16}\\ \hline	
		\end{tabular}
\end{table}

\section{Smart Buildings (Intelligente Gebäude)\label{sec:smart_buildings} }
\begin{figure}[!htb]
				\centering
				\includegraphics[width=\textwidth]{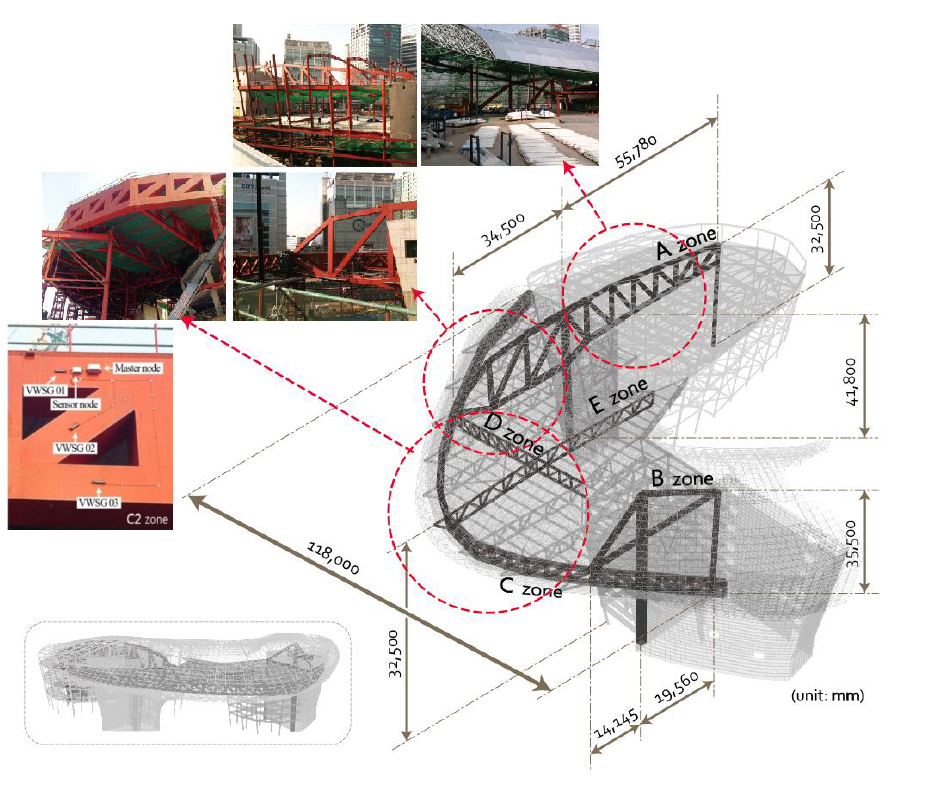}
				\caption{Beispiel für ein Echtzeitmonitoring-System für das Design Plaza Building in Seoul/Korea, Bildquelle: modifiziert nach \cite{Park13}}
				 \label{fig:dpb}
		\end{figure}

Auch Gebäude können große Datenmengen generieren. Je nach Anwendungsszenario sind hier unterschiedliche Fälle denkbar:
\begin{enumerate}
\item In vielen Gebäuden werden zunehmend Sensoren installiert, die Nutzungsdaten wie Energieverbrauch, Innentemperatur, Heizleistung, Wasserverbrauch, lokale Wetterdaten (z.B. Wind) usw. erfassen. Hier besteht ein Trend zu einer zunehmenden zeitlichen und örtlichen Auflösung. Während früher Zählerdaten in sehr großen Zeitabständen (z.B. einmal pro Jahr) erfasst wurden, sind inzwischen viele diese Daten mit Abtastzeiten von 15 min verfügbar (siehe z.B. \cite{Hammerstrom07,Kavousian13,Paetz11,Waczowicz14_Eng}). Außerdem werden Sensordaten gespeichert, die früher nur in lokalen Regelkreisen (z.B. Temperaturmessung in einem Heizungsthermostat) verwendet wurden. Ein Ziel bei der Nutzung solcher Daten besteht darin, dem Nutzer ein Feedback über seine Verbrauchsdaten zu geben und so zur effizienten Nutzung zu motivieren \cite{Paetz11,VanDam10}. Darüber hinaus sind solche Daten auch für die Wartung und Fehlerdetektion interessant, weil sie atypische Verläufe anzeigen können, die z.B. auf kurzzeitige Ereignisse wie Sensorfehler, offene Fenster und abgedeckte Lüftungen \cite{Wijayasekara14} oder langfristig auftretende Effekte wie fehlerhafte Wärmeisolierungen hinweisen können.  

Um typische Verbrauchsmuster aufzuzeigen ist es sinnvoll, Daten zu visualisieren oder Nutzersubgruppen mit Clusterverfahren zu identifizieren. Entsprechende Beispiele für die Visualisierung und Clusteranalysen zeigt \refFig{fig:peninsula}. 
\begin{figure}[p]
				\centering
				\includegraphics[width=0.58\textwidth]{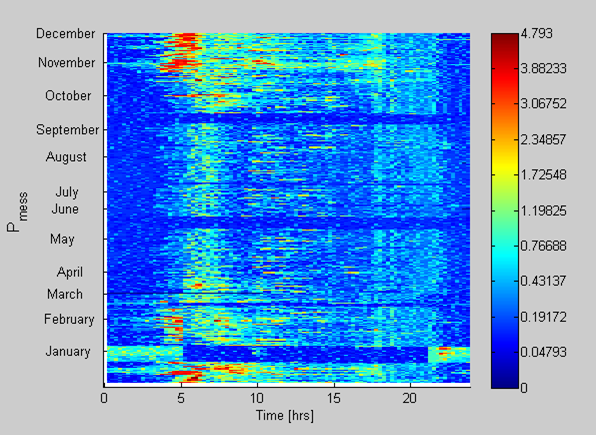}
				\includegraphics[width=0.58\textwidth]{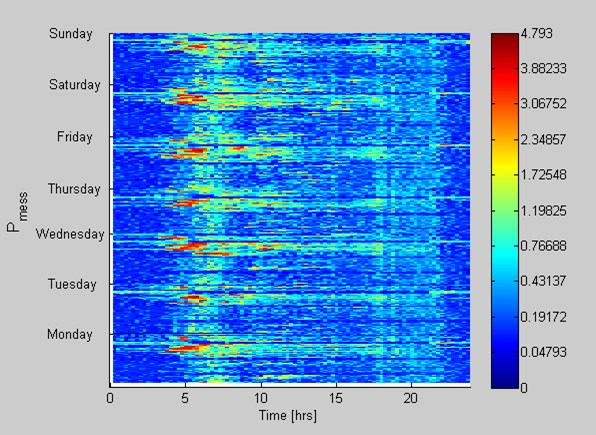}
				\includegraphics[width=0.58\textwidth]{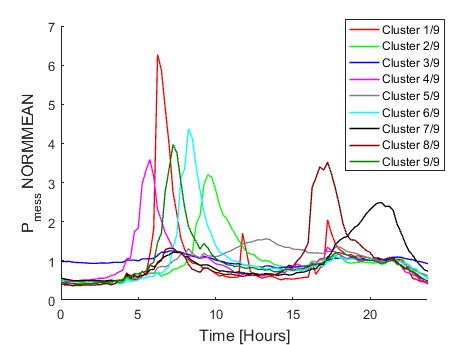}
				\caption{Analyse des Energieverbrauchs von Daten aus dem Olympic Peninsula Datensatz \cite{Hammerstrom07}: 
				a. Visualisierung eines Beispielkunden (ID 870) als Heatmap mit farbig kodiertem Energieverbrauch, Tageskurven über Monate sortiert  (oben), \newline
				b. wie a., aber über Wochentage sortiert, \newline
				c. Clusteranalyse mit 9 Clustern der normierten Energieverbrauchszeitreihen für 21894 Tageskurven verschiedener Kunden zum Identifizieren typischer Verbrauchsmuster 
				\cite{Waczowicz14_Eng}}
				 \label{fig:peninsula}
		\end{figure}
    
\item Viele Kommunen oder Betreiber von Gebäuden (z.B. in Industriegewerbeparks) haben inzwischen umfangreiche Daten zur Beschreibung ihres Gebäudebestands, siehe z.B. \cite{Kim12}. Beispiele hierfür gehen von relativ einfachen Merkmalen (z.B. Baujahr, Grundflächen, Etagenanzahl) über umfangreiche CAD-Daten (z.B. Lage Fenster, Wanddicken). Hierfür gibt es Standards für Building Information Models, um Daten aus unterschiedlichen Systemen zusammenführen zu können, siehe z.B. \cite{Cerovsek11,Geiger15,Koenig15,Loewner12}. Diese Daten können dann mit Daten über die Gebäudenutzung assoziiert werden (z.B. Energieverbrauchszeitreihen in unterschiedlicher zeitlicher Auflösung, siehe 1.).  Solche Daten können dann genutzt werden, um Simulationsmodelle für existierende und geplante Gebäudebestände aufzubauen, siehe z.B. \cite{Lauster13}. 
\item Zusätzlich können in die Grundstruktur von Bauwerken Sensoren (z.B. Beschleunigungsmesser, Dehnmessstreifen) fest integriert werden, entweder für den Verbleib über die gesamte Lebenszeit von Gebäuden oder nur während bestimmter Phasen wie in der Bauphase. Diese Ansätze werden unter dem Namen Structural Health Monitoring (SHM) zusammengefasst \cite{Balageas06,Dragos16}. Gebäude können hier z.B. große Hallen mit Tragwerkskonstruktionen \cite{Park13} oder Brücken \cite{Jang10,Rice10} sein. Die Sensoren müssen dazu entweder fest verkabelt werden oder über ein drahtloses System Informationen übermitteln \cite{Park13,Torfs13}.  Solche Sensoren können dazu verwendet werden, unerwünschte Änderungen (z.B. Verformungen, zunehmende Schwingungen) zu erkennen. Ein Beispiel für ein solches System zeigt \refFig{fig:dpb}. Hier sind verschiedene Ziele denkbar, vom Langzeitmonitoring des allgemeinen Gebäudezustandes mit dem Ziel einer Schadensprognose \cite{Farrar07} oder spezifischere Untersuchungen wie eine Korrosionsüberwachung mit mobilen oder fest installierten Messungen \cite{Holst07} bis hin zur Schadenserkennung nach Erdbeben \cite{Jiang07,Todorovska10}. Aus Sicht von Datenanalysen können entweder Zeitreihen auf Trends untersucht werden oder nach Ausreißern in einem Datensatz gesucht werden, der das normale Verhalten eines Bauwerks widerspiegelt, siehe z.B. \cite{Gul09}.
\end{enumerate}

\section{Smart Grids (Intelligente Netze)\label{sec:smart_grids}}
Eine weitere Erhöhung der Datenmengen resultiert aus der Vernetzung vieler datengenerierender und intelligenter Komponenten zu einem Gesamtsystem. Hier wird in der Regel der Begriff Smart Grids verwendet. 

Besonders populär ist dieser Begriff für das Energiesystem, weil Smart Grids als eine Schlüsseltechnologie zur Beherrschung von Systemen mit vielen lokal verteilten und fluktuierenden Einspeisungen (z.B. Solar- und Windkraftwerke) gesehen werden \cite{Fang12}. In Deutschland wird im Zuge der ''Energiewende'' ein Anteil von 80\% erneuerbarer Energiequellen zur Stromerzeugung angestrebt \cite{Scholz14}. Ein solches System muss dann durch verschiedene Maßnahmen, wie 
\begin{itemize}
\item die Integration von Strom-, Gas-, Wärme- und Dampfnetzen inkl. der zugehörigen Datenströme,
\item den Betrieb von verteilten Speichern, 
\item die Anpassung des Verbrauches an die jeweilige Erzeugungs- und Netzsituation (Demand Response) und 
\item ein geeignetes Marktdesign und geeignete regulatorische Vorgaben
\end{itemize}
stabilisiert werden. Jeder dieser Schritte erfordert das Aggregieren von Sensorinformationen, Modellierungen und Visualisierungen zum besseren Problemverständnis sowie das automatische oder zumindest halbautomatische Erstellen von Prognosemodellen und deren Nutzung in Optimierungsverfahren. Hierzu laufen vielfältige Forschungsarbeiten, z.B. in Feldversuchen \cite{Hofmann15,Witzsch15} und im Energy Lab 2.0 des KIT \cite{Hagenmeyer16}.  

Auch hier können intelligente Gebäude eine zentrale Rolle spielen. Beispiele hierfür sind Strategien zum Demand Response, die  Energieverbraucher (z.B. Wärmepumpen, Klimaanlagen usw.) netzdienlich einsetzen, siehe z.B. \cite{Koutitas12,Paetz11}.  

\section{Zusammenfassung}
In vielen Anwendungsgebieten haben in den letzten Jahren die gesammelten Datenmengen deutlich zugenommen. Damit besteht die Chance, dass in diesen Daten interessante Informationen enthalten sind, aus denen nützliche Erkenntnisse generiert werden können. Das setzt eine enge Zusammenarbeit von Anwendungs- und Datenanalyseexperten voraus. Prinzipiell stehen etablierte Analysemethoden bereit. Bei sehr großen Datenmengen (''Big Data'') ist speziell darauf zu achten, geeignete Methoden und Tools zu verwenden, die die gewünschten Ergebnisse bei akzeptablem Aufwand selbst bei zusätzlichen Schwierigkeiten wie Unsicherheiten und schlechter Datenqualität liefern. Hier besteht weiterer Forschungsbedarf, um mit den zunehmend anspruchsvolleren Anwendungen Schritt zu halten. 

Der Beitrag diskutiert ausgewählte Anwendungsgebiete wie Smart Buildings und Smart Grids, in denen erste Big-Data-Aufgaben erfolgreich gelöst wurden und weitere Aufgaben vielversprechend erscheinen. Das Ziel des Beitrags bestand darin, die Sicht der Datenanalyse zu erläutern, um die Diskussion über zukünftige Aufgaben mit Experten in den jeweiligen Anwendungen anzuregen. 

{\bf Danksagung}: Diese Arbeit wurde von der Helmholtz-Gemeinschaft innerhalb der gemeinsamen Initiative  ''Energie System 2050 - Ein Beitrag des Forschungsbereichs Energie'' unterstützt. 

%\bibliographystyle{abbrv}
%\bibliography{e:/mytex/biblio/biosignal,e:/mytex/biblio/zusatzeintraege}

%	
%%%%%%%%%%%%%%%%%%%%%%%%%%%%%%%%%%%%%%%%%%%%%%%%%%%%%%%%%%%%%%%%%%%%%%
% literature
%%%%%%%%%%%%%%%%%%%%%%%%%%%%%%%%%%%%%%%%%%%%%%%%%%%%%%%%%%%%%%%%%%%%%%

%%%%%%%%%%%%%%%%%%%%%%%%%%%%%%%%%%%%%%%%%%%%%%%%%%%%%%%%%%%%%%%%%%%%%%%%%%%%%%%%%%%%%%%%%%%%%%%%%%%%%%%%%%%%%%%%%%%%%%%%
\end{document}